\documentclass[letterpaper]{article}
\usepackage{iccc}
\usepackage{times}
\usepackage{helvet}
\usepackage{courier}
\usepackage{graphicx}
\usepackage{hyperref}
\usepackage{multirow}
\usepackage{xcolor}

\newcommand{\ie}{i.\,e., }
\newcommand{\eg}{e.\,g., }
\title{An approach to Beethoven's 10th Symphony}
\author{Paula Mu\~{n}oz Lago$^1$, Gonzalo M\'endez$^{1,2}$\\
$^1$Facultad de Informática\\
$^2$Instituto de Tecnología del Conocimiento\\
Universidad Complutense de Madrid\\
Madrid, Espa\~{n}a\\
pmunoz06@ucm.es, gmendez@fdi.ucm.es\\
}
\begin{document} 
\maketitle
\begin{abstract}
\begin{quote}
Ludwig van Beethoven composed his symphonies between 1799 and 1825, when he was writing his Tenth symphony. As we dispose of a great amount of data belonging to his work, the purpose of this paper is to investigate the possibility of extracting patterns on his compositional model from symbolic data and generate what would have been his last symphony, the Tenth. A neural network model has been built based on the Long Short-Therm Memory (LSTM) neural networks. After training the model, the generated music has been analysed by comparing the input data with the results, and establishing differences between the generated outputs based on the training data used to obtain them. 
The structure of the outputs strongly depends on the symphonies used to train the network.

\end{quote}
\end{abstract}

\section{Introduction} 
Romantic composer Ludwig van Beethoven wrote his Symphonies from 1799 to 1824, when he finished the No. 9 \cite{cooper2000beethoven}. Although there is no constancy of the existence of the 10th Symphony score, there exists some manuscripts found in Beethoven's house after his death that are thought to be part of the upcoming Symphony. In 1988 Barry Cooper tried to finish it, building from 50 of those fragments the first movement of the Symphony. Those manuscripts are kept in the museum dedicated to his life in his natal city, Bonn, although they can be seen online \footnote{\url{https://bit.ly/2BKPAOx}}. The public manuscript is not easy to read and understand, so that existing data will not be used in this paper. 

The goal of this work is to generate music, based on Beethoven's compositional model, obtaining the conductor's score with all the orchestra instrument's parts. Two approaches to the goal were established; 
firstly, the system was trained with each instrument individually, to generate all the different instrument's parts, and then put them all together in a conductor's score.
Nevertheless training each part individually lead to a lack of coordination, so a new approach was addressed.
This second approach consisted in training the system with information from different instruments at the same time, extracting the data \textit{vertically}, to maintain the harmony (\textit{vertical}) and time (\textit{horizontal}) information.

\begin{figure}[t]
	\centering
	\includegraphics[width = 0.35\textwidth]{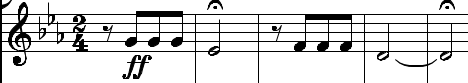}
	\caption{Snippet of Beethoven's Fifth Symphony in C minor}
	\label{fig:sinippet5}
\end{figure}

The output is intended to be fully dependent of the system prediction. The only forced characteristics' are the symphony's tempo and key, as the sheets found in Beethoven's house had 3 flats (\ie Eb key, also called C minor), and the measure was a 6/8. That key had a great significance for Beethoven \cite{cooper2000beethoven}, as it is said that it represent a ``stormy and heroic tonality'', and it is used in works of unusual intensity, such as the Fifth Symphony (Figure \ref{fig:sinippet5}).


This paper is structured as follows. Previous work on using Artificial Intelligence in computational creativity and concretely in music generation is exposed in the State of the art section. After that, musical definitions needed to understand requirements, limitations and characteristics found in the results are introduced. Later, the work developed for this paper is explained in detail, presenting the Deep Learning technique used, the needed toolkits, and how the data was represented. Then, the Music generation subsection is divided in: dataset creation, training, and prediction. The results section is focused on explaining the reason why the system returns a certain output when trained with a specific set of symphonies. The conclusions and future work are described in the last sections.

\section{State of the art}
Studied since the latter half of the 20th century, Computational Creativity can still be considered as a novel field. 
There has been relevant experiments on Linguistic creativity, such as narrative \cite{Gervas}, getting to write the lines of a musical called \textit{Beyond the fence} \cite{beyondTheFence}, poem \cite{montfort} or jokes generation \cite{ritchie2009can}. 
Visual Arts creativity has also received attention. 
Work on this field range from AARON, \cite{CohenPaints}, a robot capable of painting taking a brush with it is robotic arm, or the \textit{Painting Fool} \cite{colton2012painting}, software intended to be taken as a real artist which emulates several styles, to the algorithm designed by \textit{Obvious}\footnote{\url{http://obvious-art.com/index.html}} that generated the \textit{Portrait of Edmond de Belamy (2018)}, getting the price of \$432.500. 

\subsection{Music creativity}
This Computational Creativity sub-field started in the early 50's, and its most relevant works are mainly focused on generating coherent sounds and scores for the human musicians use. Techniques used to broad this research are Markov chains, Recurrent Neural Networks, Long short term memory neural network or genetic algorithms.

The first Artificial Intelligence technique used for this purpose were the Markov chains. This model defines the probability for an event to happen based on previous ones, storing them in a transition matrix. An example of the application of the Markov Chains is \textit{ILLIAC} \cite{hiller1958musical}. This machine generated the \textit{ILLIAC's suite}, a string quartet. Generated notes were tested by heuristic compositional rules. In case that the rules were not violated, they were kept, otherwise a backtracking process was followed. This project excluded any emotional or expressive generation, by just focusing on the notes. 
Later on, a system called \textit{CHORAL}, which produced the corresponding harmonisation of a given Bach Choral, was developed creating rules and setting heuristics in a logic-programming language created by the author for this purpose \cite{BachLogic}.

Since music is built on themes and motifs repeating over time, it makes Long Short Term Memory (LSTM) neural networks a reasonable option for computer music creativity. This method incorporates the ability to learn long-term dependencies, by improving the cells or neurons in the Recursive Neural Network (RNN) graph.
Melodies generated with LSTM networks have resulted more musically plausible than other models' results, such as Gated Recurrent Unit (GNR) \cite{nayebi2015gruv}. The first music generation project that used neural networks is \textit{MUSACT} \cite{bharucha1992musact}, which focuses on learning the harmonic model and generates expectations after listening to a certain chord. Another example is \textit{BachBot} \cite{liang2017automatic}, which composes and completes music in the style of Bach chorales using an LSTM generative model. They conducted a discrimination test to determine if the generated music was similar to Bach's chorales with 2336 participants, getting a rate of only a 1\%  of the people correctly determining which music was generated with \textit{BachBot}.

Nevertheless, other techniques have resulted useful in the music generation task too. \textit{EMI}, \cite{cope1996experiments} has successfully emulated Mozart, Brahams, Bach, Rachmaninoff or Chopin's music, generating new music. It searches a pattern, in at least two existing pieces of a concrete compositor. Using one of the artist scores, it locates the signatures and composes music between them, by using a rule analyser.
 \textit{IAMUS} \cite{Quintana2013Iamus} is capable of composing a full score in 8 minutes, using genetic algorithms. Its music has been played by the London Symphony Orchestra. In this case, chromosomes including all the notes information are randomly generated, and fitness functions are applied to each of them. If a note is codified to be played by a violin and this instrument does not have the possibility to play that note, it is changed. After generating around 100 scores, a human composer chooses the best one as the final output.

Another challenging field relating Musical Creativity is Music Improvisation, since it has more difficulties from a creative point of view. Using Genetic algorithms, GenJam \cite{GenJam} emulates a Jazz musician in his or her improvisation learning process. \textit{Continuator} \cite{pachet2003continuator} uses a Markov model to generate music in standalone mode, as continuations of musician’s input, or as interactive improvisation.

\section{Musical definitions}
In order to fully understand the paper development it is relevant the knowledge on the following musical definitions.

\begin{itemize}
	\item Note: Musical event that describes a sound. Besides the note name, it contains information about the duration, or the pitch class (Figure \ref{fig:notes}) \cite{note}.
	\begin{figure}[t]
		\centering
		\includegraphics[width = 0.5\textwidth]{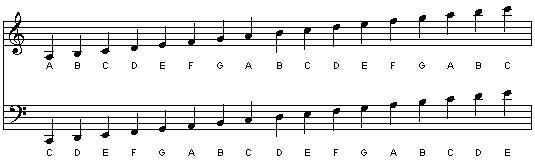}
		\caption{Note names of the chromatic scale}
		\label{fig:notes}
	\end{figure}
	\item Pitch: Property of sounds that allows a frequency scale ordering, fixing the position in the scale by distinguishing between "higher" or "lower" sounds \cite{pitch}.
	\item Clef: Musical symbol used to determine the name and pitch of the written notes, it is the first symbol that appears in the score in Wenstern's notation. The tree types are: F (second stave from Figure \ref{fig:notes}), C and G (first stave from Figure \ref{fig:notes}) \cite{clef}.
	\item Key signature: Group of sharp or flat symbols placed after the clef of after a double bar in the course of the composition, it determines the notes that will be altered from their natural pitch. A sharp raises one semitone the natural note, while the flat lowers it \cite{keysignature}. 
	\item Time signature: It determines the metre of the piece. Appears next to the key signature or in the course of the composition. As we can see in the example of Figure \ref{fig:sinippet5}, the 2/4 time signature means that there are two crotchets per beat \cite{timesignature}.
	\item Dynamics: Refers to the volume in which notes and sounds are expressed. It is symbolically encoded with marks, whose main levels are \textit{p} (\ie piano) or \textit{f} (\ie forte) \cite{dynamics}.
	\item Harmony: Organisation of simultaneous sounds along the time axis. \cite{MusicDataMining}
\end{itemize}

\section{Work description}
\subsection{Technical background}
\subsubsection{Deep learning: LSTM Networks}
Included in the field of Machine Learning, Deep Learning involves the use of artificial neural networks \cite{gulli2017deep}. There exists several types of neural networks, such as Deep Neural, Deep Brief and Recurrent Neural Networks (RNN). In this paper we work with the last ones, since we need to process sequential data, assuming that each event depends on previous ones. The most accurate RNN variant is the LSTM. As proved with Figure \ref{fig:sinippet5}, we need the memory that this type of networks own. On it we find the sequence F - F - F, a predictor without memory would return another F, although by learning from the notes before, it can extract that after three equal notes, it is probable that the upcoming note is two tones below the last one.

Proposed in 1997, LSTM neural networks can learn long-term dependencies, improving the cells or neurons in the RNN graph. They have the ability to connect previous knowledge to a present task. Each cell has memory, and it decides to store or forget a data based on a given priority (\ie represented as weights), assigned by the algorithm after the learning process.
\begin{figure}[t]
	\centering
	\includegraphics[width = 0.4\textwidth]{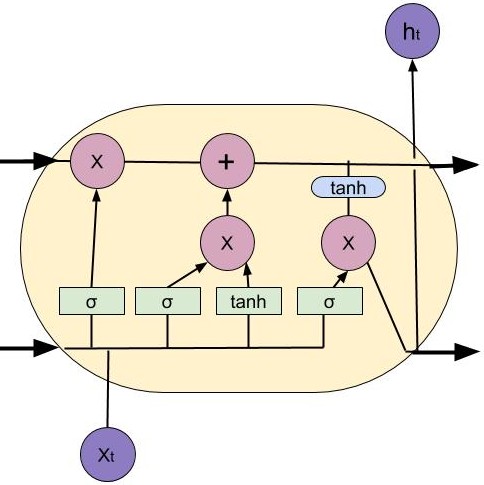}
	\caption{LSTM neural network cell}
	\label{fig:LSTM}
\end{figure}
Figure \ref{fig:LSTM} shows a LSTM cell or neuron. The top line represents the flow of the cell state, which can be altered up to three times. 
The first layer, sigmoid ($\sigma$),  takes information from the previous state and determines if it is useful or not, returning a value between 0 and 1. As it is shown with the vertical arrow, it directly affects to the flow of the cell state. 
The second layer is composed of the combination of the sigmoid ($\sigma$) and \textit{tanh} functions,  which chooses the data to be updated from the previous state, and creates a vector of candidate values to be added to the current cell state.
The final sigmoid ($\sigma$) layer determines the output, by deciding which parts of the state are more relevant. Those will be combined with a \textit{tanh} function, converting the current state into values between 1 and -1 \cite{gulli2017deep}.

\subsection{Toolkits}
This project has been developed in Python. Data extracting and processing from the scores has been performed using the python's library Music21 \cite{cuthbert2010music21}, which allows parsing and generating scores in different formats. Furthermore, every musical action and representation that we needed to perform, was made possible using that library. 
For the Deep Learning engine we have used Keras \footnote{\url{https://keras.io/}} \cite{gulli2017deep}. 
Finally, in order to manage the score formats, Musescore \footnote{\url{https://musescore.com/}} (\ie open source program available for every platform) brought us the possibility to import and export the symphonies, so we could see the score and listen to it at the same time.

\subsubsection{Data representation}
Several ways of representing the Beethoven Symphonies' scores have been studied for this paper. 
\begin{figure}[t]
	\centering
	\includegraphics[width = 0.47\textwidth]{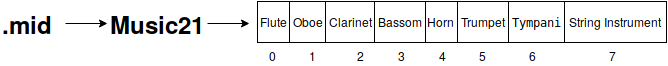}
	\caption{Music21 MIDI files parsing}
	\label{fig:MIDI}
\end{figure}
Firstly, we used  MIDI files as an input for our system, as it is a data file which contains information about the sounds: what note is played, when, and how long or loud. Figure \ref{fig:MIDI} shows the problem we boarded with the MIDI input. Music21 was unable to differentiate between the different string instrument's MIDI channels.

 Since the main goal of this paper is to obtain a score including every orchestra instrument's part, the input files format were changed to symbolic data in mxl. This extension refers to a compressed music score, which Music21 easily processes. Mxl files are the compressed format of the so called MusicXML \cite{good2001musicxml}.
 In order to represent the output of the training, i.e. the weights of the different notes and durations, the model also returns a HDF5 file, \ie \textit{Hierarchical Data Format} version 5, commonly used to store big quantities of data.
 After the prediction process, given the obtained weights, Music21 allows us to generate the final output in MIDI or MusicXML, formats accepted by Musescore (so the score can be visualised and played).
 
\subsection{Music generation}
In this paper, we have established two different approaches in order to obtain the expected result, which is the new Beethoven's Tenth Symphony. 

The first approach is based on generating all the different orchestra instrument's parts individually. 
By training each instrument with a concrete existing set of symphonies, we have obtained each part. After that, we have manually joined all the different instruments to study if the overall symphony was musically valid. Since each instrument was trained without information of the other instruments, the obtained conductor score presented a lack of coordination between them.
  
The second approach was intended to increase the coordination between each instrument, by combining time  (\ie \textit{horizontal}) and harmony (\ie \textit{vertically}) information. This was achieved by changing the data extraction phase from obtaining separately each complete instrument part to obtain every instrument part at every beat. This allowed us to train a set of instruments at the same time from a concrete set of symphonies. 
This way, the generated parts present a considerable increment of coordination and it is easier to differentiate each musical phrase, as each instrument part respects or accompanies the others.

\subsubsection{Dataset creation}
\label{dataset}
As previously mentioned, all the Beethoven symphonies have been converted to an mxl file, which constitutes the dataset or corpus that we have used to obtain the desired results. Also, the instrument or instruments with which the system  works has to be established, so the Python module music21 can divide the mxl score into all the present instruments, and take only the desired parts. This way, in the first approach, where the goal is to obtain each instrument's part individually, the note names and durations are stored in an independent file, being the different tuples of note names and durations the training data. Nevertheless, as the second approach trains the model with a set of chosen instruments at the same time, we need to store, besides the note name and duration, the offset (\ie time data relating to the moment in which the note is being played regarding the score) and the name of the instrument that plays it. The offset information will be used to sort the data. After making sure that the events are sorted in a time-line as they are in the original score, it can be removed from the dataset, in order to reduce the data dimensionality and to avoid an overlearning problem in the model. This way, the training data will be composed of the different tuples of note names, note durations and the instrument's name playing it (introduced in the second approach). At this point, a dictionary to encode each data tuple as a number is created, so the neural network can work with it. This dictionary will be also used for the decoding phase, after the prediction.

\subsubsection{Training}
Finally, we can generate the training data (\ie sequence input and output). By establishing a certain sequence length, the output for each input sequence will be the first note that comes after that sequence. 
\begin{figure}[t]
	\centering
	\includegraphics[width = 0.47\textwidth]{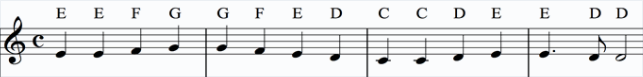}
	\caption{Violin's \textit{Ode To Joy} snippet}
	\label{fig:OdeToJoy}
\end{figure}
\begin{table}[h]
	\centering
	\begin{tabular}{cc}
		\textbf{sequence\_in} & \textbf{sequence\_out} \\
		{[}(E, 1), (E, 1){]}  & {[}(F, 1){]}           \\
		{[}(E, 1), (F, 1){]}  & {[}(G, 1){]}           \\
		{[}(F, 1), (G, 1){]}  & {[}(G, 1){]}           \\
		{[}(G, 1), (G, 1){]}  & {[}(F, 1){]}          
	\end{tabular}
	\caption{Figure's \ref{fig:OdeToJoy} input and output sequences using a length = 2}
	\label{table:Sequences}
\end{table}

For example, setting a sequence length equal to two, the first stage of the system's work flow (\ie data extraction) for the Figure \ref{fig:OdeToJoy} input would be the shown in Table \ref{table:Sequences}. It is important to take into account that in case of establishing a big sequence length, the model may generalise, while setting a small sequence length may lead to an  overlearning problem.

In case of the input, reshaping into a 3 dimension matrix is needed so it is compatible with the LSTM layers, using Python's numpy module. The first dimension or shape of the network is the number of unique different sequences (\ie \textit{sequence\_in} in Table \ref{table:Sequences}) obtained in the last step, the second one is the previously established sequence length and finally the last dimension is forced to be 1, so it has just one input information per sequence length. After that, the software normalises the input into sequential values, from 0 to 1. In case of the output, it is converted into a categorical model.

The next step is to create the model, which follows a stacked LSTM architecture, since the larger the depth, the less neurons per layer the network needs, and it is faster \cite{DBLP}. There's no formula established to determine how many layers the network should have, and how many neurons would work better for each layer, so one of the tasks during the development of this project has been to obtain that information empirically.

\begin{figure}[t]
	\centering
	\includegraphics[width = 0.48\textwidth]{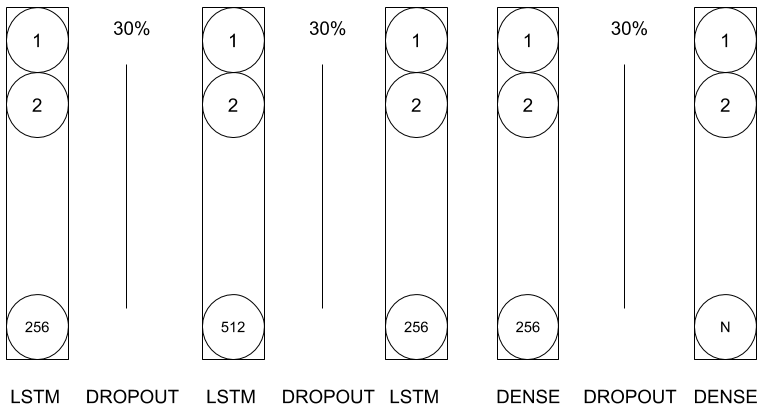}
	\caption{Model, being N the number of different tuples (note name, duration)}
	\label{fig:MODEL}
\end{figure}

Our final network is composed 3 different types of layers. The most relevant ones are the LSTM layers, which take the sequences and return new ones. Then, the Dropout layers prevent overfitting, ignoring randomly selected neurons during the training, setting those inputs to 0. The Dense (Density) layer serves as a full connection mechanism. This layer is the last one, so the system returns the same number of outputs as the different numbers of tuples (note name, note duration) the input data had. Finally, the activation function used for every layer is set, and it determines how each node's output is represented. In this case, the softmax function (\ie linear activation) is used, allowing the output to be interpreted as a probability between 0 and 1.

Once the model is built and the input and output data are ready, it gets trained, generating an hdf5 file containing the weights (\ie input notes' priorities).

\subsubsection{Prediction}
\label{prediction}
For this task, the network input is generated again, as in the previous process (see Table \ref{table:Sequences}). Since it needs to work over the same model, it is created again, with the same parameters, but now, instead of training the model, it loads the generated weights from the previous process (\ie the hdf5 file). It is important at this point that the network input shapes and the loaded weights have the same dimensions.
Once the model is ready, the encoding dictionary built during the dataset creation phase is inverted, for decoding the prediction results.

Then, a random sequence from the input is extracted and the trained model starts to predict notes till it gets to a desired time duration. As in the training, this random sequence has to be reshaped into a 3 dimension matrix. The first dimension corresponds to the number of unique sequences, the second to the length of the sequence and the third, as in the training, is forced to be 1. 
The output of the prediction is an array with a probability for each tuple. Then, the system sorts the values from the greatest probability to the lowest. As previously mentioned, if we are trying to generate a single instrument score, the tuple is composed of (note name, note duration), while if we want to generate a conductor's score composed of several instruments, the instrument identification has to be included in the tuple.

\begin{figure}[t]
	\centering
	\includegraphics[width = 0.47\textwidth]{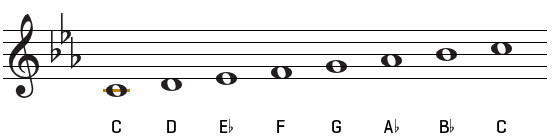}
	\caption{Eb Key scale}
	\label{fig:Scale}
\end{figure}

Once it has the indexes of the most probable notes, the system can work on the given tuples accessing to the decoding matrix. It forces the predicted notes to have a duration greater or equal to 0.5 (quaver), for the score's simplicity. Another important restriction is to give priority to notes that belong to the key scale used in the new score (\ie Eb, due to conserved manuscripts present that key). Notes belonging to that key are present in Figure \ref{fig:Scale}.

Some other restrictions performed during the prediction phase are: octave reduction, and rest management. Firstly, if the note with highest prediction differs more than one octave from the last one generated, it is performed and octave reduction in order to get closer to the previous one, but not modifying the note predicted, since it will not be easy to play for a musician. 
Another change made at this point is that if the predicted note and the previous one are rests, the lengths are added. This can only be applied to rests since we need to have several identical notes following (see Figure \ref{fig:sinippet5}).
After choosing the most appropriate note, the index of the selected note is added to the pattern, which serves as an input for the next prediction.

Once the system has all the required predicted information (notes, chords, rests, and all the needed information such as their durations or the instrument that plays them) it is processed, so finally a MusicXML and MIDI files are created using Music21. In this problem, since it is all about creativity, we do not have an automatic validation step, present in the majority of machine learning problems, due to the nonexistence of a correct solution. Nevertheless, a human validation has been used to evaluate the resulting score, modifying the model or prediction restriction rules based on the evaluation feedback.

\subsection{Results} 
The system output differs from the information given to the training, although once with the same trained data, the system predicts the same score, which denotes a lack of variability. 
\subsubsection{Approach 1: Generating individual instruments melodies}
\begin{figure}[t]
	\centering
	\includegraphics[width = 0.45\textwidth]{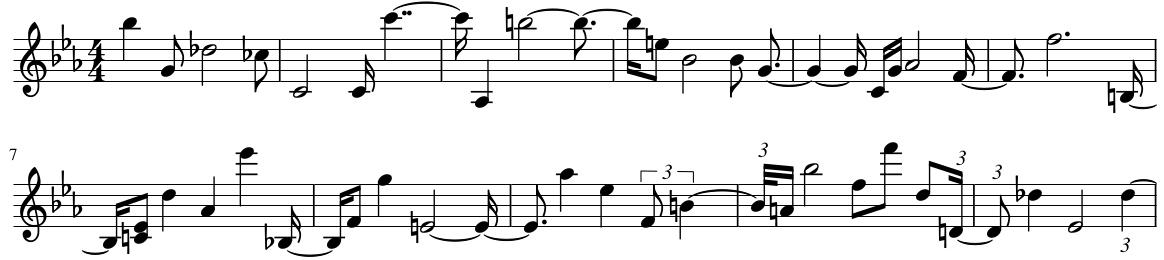}
	\caption{Results from training with the Fifth Symphony}
	\label{fig:FifthDurations}
\end{figure}

The first experiment consisted on generating music based on the Violin I parts of every Fifth Symphony's movements, the first three staves of the output obtained is shown in Figure \ref{fig:FifthDurations}. 

\begin{figure}[t]
	\centering
	\includegraphics[width = 0.45\textwidth]{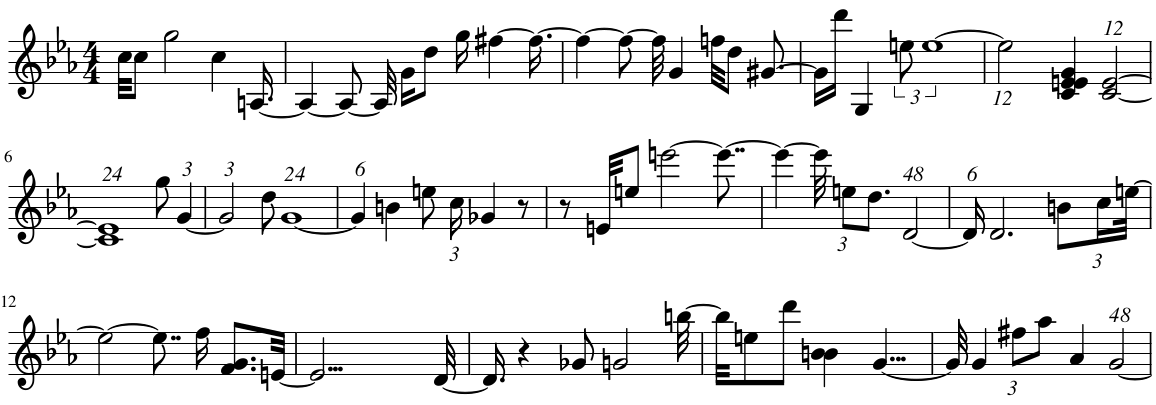}
	\caption{Results from training with the Fifth Symphony allowing rests}
	\label{fig:FifthRests}
\end{figure}

It can be seen that different measures showed up, such as quarter, eight, sixteenth or half notes but also thirty-second notes, and a motif shows up. In the first two staves, a half note appears tied to an eight and a sixteenth note in several compass. However, there are no rests, so the next step at this point was to retrain the system, again with the most famous symphony, but allowing rests to appear. The results can be seen in Figure \ref{fig:FifthRests}.
Again, although a different score is generated, we can distinguish some patterns in the composition.

 \begin{figure}[t!]
	\centering
	\includegraphics[width = 0.45\textwidth]{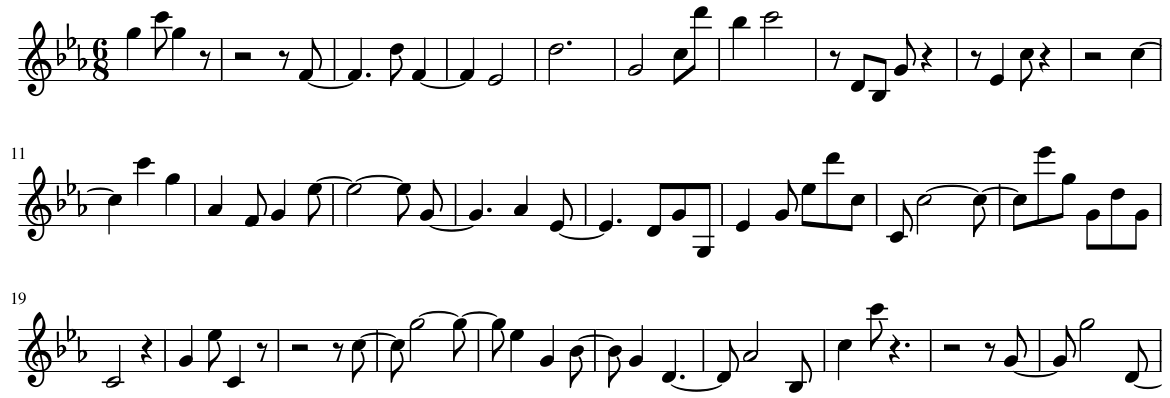}
	\caption{Results from training with the Fifth Symphony with prediction restrictions}
	\label{fig:FifthWithImprovements}
\end{figure}

At this point musical restrictions previously explained during the prediction were implemented. Time measure shown in Figures \ref{fig:FifthDurations} and \ref{fig:FifthRests} is a 4/4 as a first approach, although after discovering that Beethoven's house sketches belonging to the upcoming symphony had measure 6/8, it was set to that one. All the experiments from this point included these musical improvements. For instance, using the same weights as before, the first three staves of the outcoming score is shown in Figure \ref{fig:FifthWithImprovements}. The result differed from the obtained in the previous experiment (\ie Figure \ref{fig:FifthRests}), being the new one clearer (\eg the notes are found in a reduced range, due to the octave reduction requirement) but maintaining the motifs.

\begin{figure}[t]
	\centering
	\includegraphics[width = 0.45\textwidth]{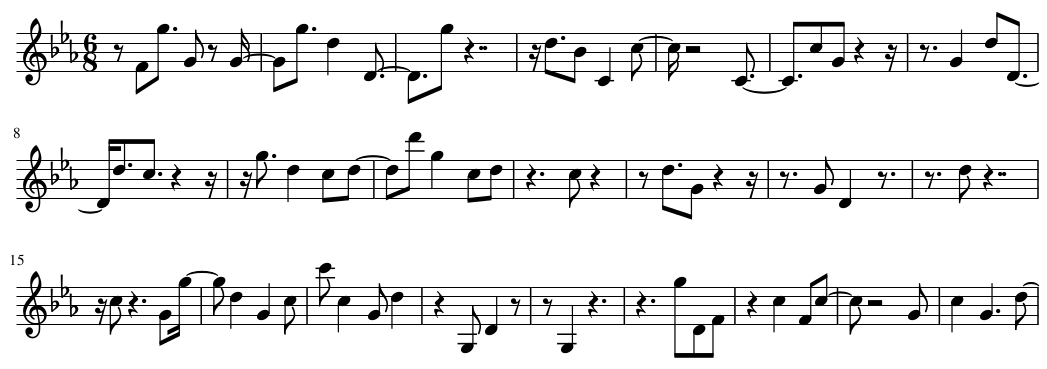}
	\caption{Results from training with the Seventh Symphony with manual improvements}
	\label{fig:Seventh}
\end{figure}

Keeping the system state, we trained it with the Seventh symphony, and generated the violins as before (Figure \ref{fig:Seventh} shows only the first three staves). The result looks similar, although we found several remarkable characteristics: the increment in the number of rests showing in the score and the lack of a pattern. 
The first characteristic may appear due to the amount of silent compasses in the second movement of this symphony \footnote{\url{http://ks4.imslp.net/files/imglnks/usimg/a/ab/IMSLP312601-PMLP01600-LvBeethoven_Symphony_No.7_BH_Werke_fs.pdf}}. Violins I start playing in compass number 50, which is not a common characteristic of the violin's part in any symphony, being usually the instrument playing the main melody. As previously mentioned, there is not an easy-to-recognise motive such as in the previous experiments. That may be caused due to the absence of a clear motive in contrast with the Fifth symphony.

 \begin{figure}[t]
	\centering
	\includegraphics[width = 0.45\textwidth]{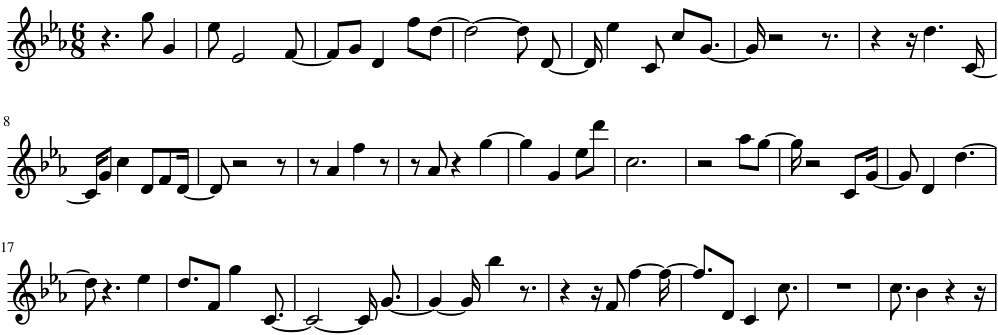}
	\caption{Results from training with the Fifth and Seventh Symphony}
	\label{fig:FifthandSeventh}
\end{figure}

\begin{figure}[t]
	\centering
	\includegraphics[width = 0.45\textwidth]{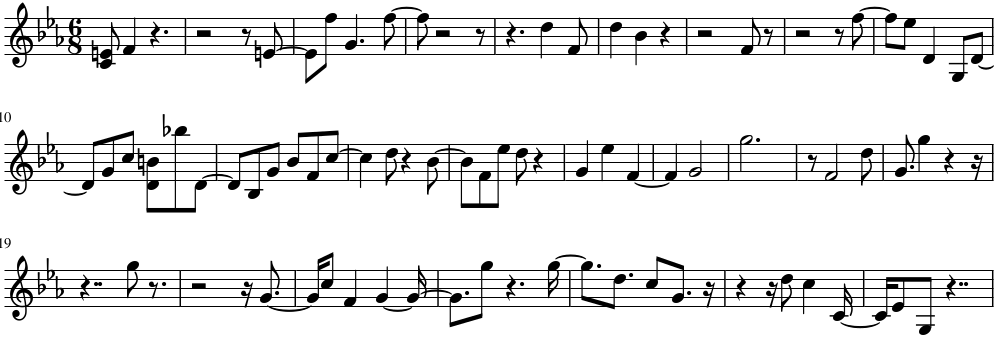}
	\caption{Results from training with the Fifth, Seventh and Ninth Symphony}
	\label{fig:FifthSeventhandNinth}
\end{figure}

After concluding with the experiments training the model with the Fifth and Seventh symphonies individually, the next step was to train the system with both of them at the same time. Figure \ref{fig:FifthandSeventh} shows the output, where it is distinguishable that the amount of rest notes was increased from other results that does not use the Seventh symphony violin's as input, but small motives present in the output obtained from training with the Fifth keeps showing. The same characteristics can be found in Figure \ref{fig:FifthSeventhandNinth}, result obtained from training the Fifth, Seventh and Ninth Symphonies' Violins.

\begin{figure}[t]
	\centering
	\includegraphics[width = 0.47\textwidth]{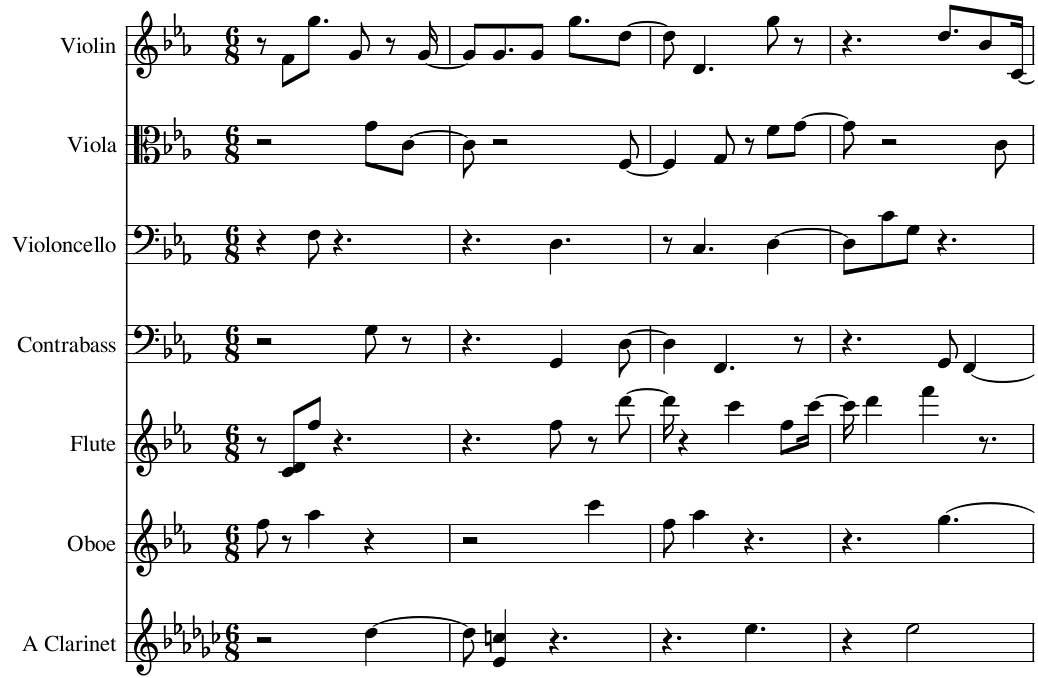}
	\caption{Results from training separately 7 different instruments with the Seventh symphony}
	\label{fig:Seventh7}
\end{figure}

After completing all the experiments previously described, the system was trained with some of the orchestra's instruments. Figure \ref{fig:Seventh7} shows the prediction result for Violin, Violas, Violoncellos, Contrabass, Flutes, Oboes and Clarinets, trained with the Seventh symphony. Although this result was obtained from training each instrument individually and putting them together manually, it is distinguishable a lack of coordination between each instrument, since each melody was generated without having knowledge on any other instrument's melody. That caused that each musical phrase from the different instruments does not coordinate with the others to generate a group sound.

\subsubsection{Approach 2: Generating  several instruments at the same time}

\begin{figure}[t]
	\centering
	\includegraphics[width = 0.47\textwidth]{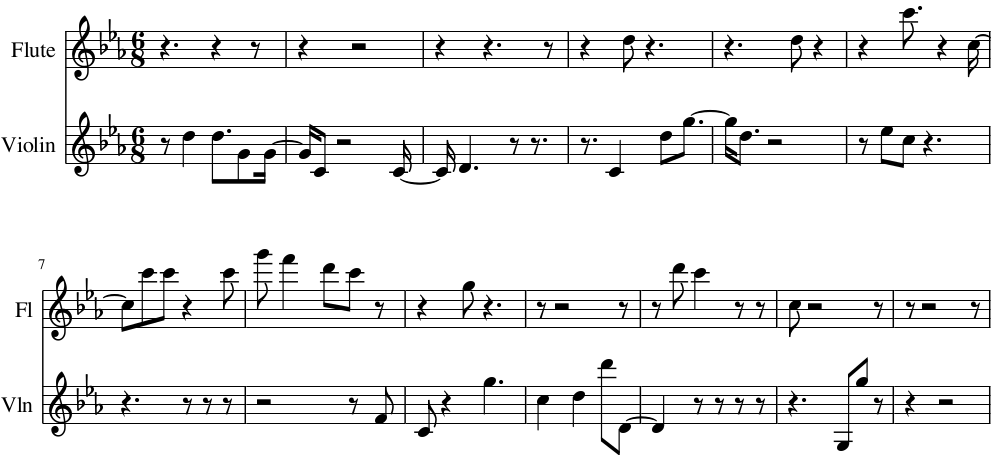}
	\caption{Second approach trained with the Seventh symphony for Flutes and Violins}
	\label{fig:SecondAproachFV}
\end{figure}

To avoid the musical disorder obtained in the previous results, the second approach was used. As explained before, in this case the system was trained with a set of desired instruments, keeping the time information but adding the harmony created between instruments, getting this way scores such as the one that is appreciated in Figure \ref{fig:SecondAproachFV}, trained with the Seventh Symphony for Violins and Flutes. This result shows how each instrument compliments the others, having the violin the main melody at the beginning, but respecting the Flute's main appearance in compasses seventh and eight, establishing a communication between them.

\begin{figure}[h]
	\centering
	\includegraphics[width = 0.47\textwidth]{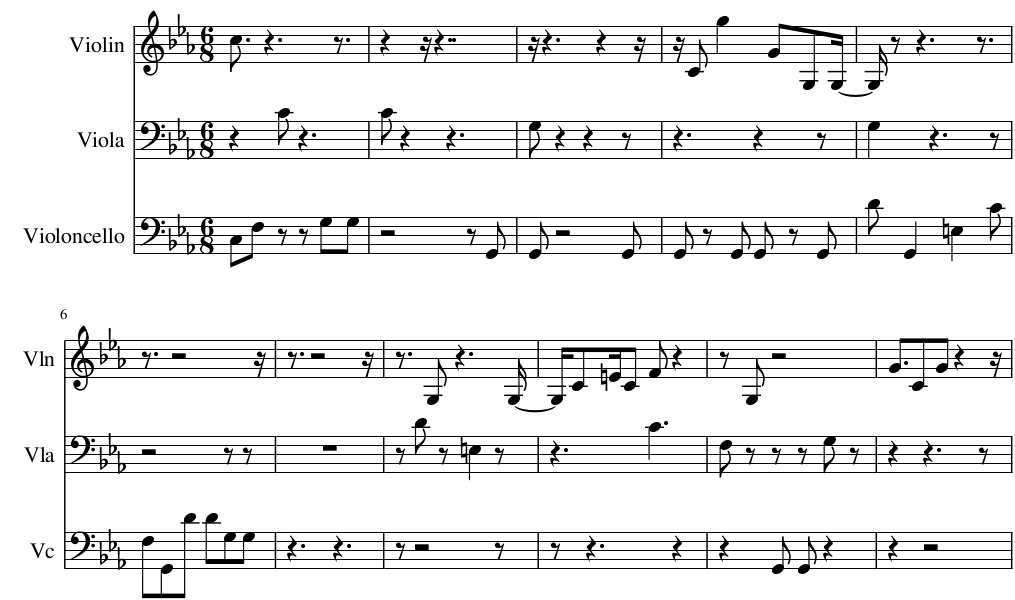}
	\caption{Score obtained from training Violins, Violas and Violoncellos with the Seventh symphony}
	\label{fig:FirstAproach}
\end{figure}

The same behaviour can be seen in the result shown in Figure \ref{fig:FirstAproach}, which shows how Violins, Violas and Violoncellos, while being trained only with the Seventh symphony, assumes a trio music by respecting the other instrument's melodies and complementing each other. It can be appreciated the differences between this score and the one shown in Figure \ref{fig:Seventh7}. In that case, the corresponding lines are the first, second and third (Violin, Viola and Violoncello). As it can be seen, the coherence of the different instruments is enhanced in the second approach.

\section{Conclusions}

\begin{table*}
	\begin{tabular}{|c|c|c|c|c|}
		\hline
		Approach                         & Instruments                                                                                                      & Symphony trained     & Details                            & Figure \\ \hline
		\multirow{7}{*}{First approach}  & \multirow{6}{*}{Violins}                                                                                         & \multirow{3}{*}{5th} & Without rests                      & 8      \\ \cline{4-5} 
		&                                                                                                                  &                      & With rests                         & 9      \\ \cline{4-5} 
		&                                                                                                                  &                      & With rests and musical restrictions & 10     \\ \cline{3-5} 
		&                                                                                                                  & 7th                  & With rests and musical restrictions & 11     \\ \cline{3-5} 
		&                                                                                                                  & 5th + 7th            & With rests and musical restrictions & 12     \\ \cline{3-5} 
		&                                                                                                                  & 5th + 7th + 9th      & With rests and musical restrictions & 13     \\ \cline{2-5} 
		& \begin{tabular}[c]{@{}c@{}}Violins, Violas, Violoncellos,\\  Contrabass, Flute, Oboe, \\ A Clarinet\end{tabular} & 7th                  & With rests and musical restrictions & 14     \\ \hline
		\multirow{2}{*}{Second approach} & Violins, Flute                                                                                                   & 7th                  & With rests and musical restrictions & 15     \\ \cline{2-5} 
		& Violins, Violas, Violoncellos                                                                                    & 7th                  & With rests and musical restrictions & 16     \\ \hline
	\end{tabular}
\caption{Summary of results}
\label{Results}
\end{table*}

This paper explores the possibility of generating new music based on Beethoven's style by a system doted with Artificial Intelligence, using LSTM neural networks, which learns and remember musical phrases of a concrete length, finally showing that it is possible to obtain music that imitates this composer's style for several instruments.

During the specification of the problem, we established two ways of approximating to the new symphony. The first one was to train and generate separately each instrument scores, and manually creating the conductor's score. The results obtained were satisfactory for each single instrument separately, getting to generate music that a musician could play, as it respects the musical standards. However, when joining all the different scores, the sound was not coordinated and the musical phrases belonging to the different instruments were not respected by the others. We concluded that with this first approach we could generate solo scores, but not group music. The second approach was intended to solve the main problem present in the first one, that the instruments were not sufficiently coordinated due to the lack of knowledge of other instrument's music while training, which is crucial in an orchestra. The solution proposed was to train and generate music belonging to different instruments at the same time. This way the results obtained were more coordinated and we could see that each instrument respected each other, having rests or accompanying the main melody when they did not have the leading voice (\eg showing a musical \textit{conversation}). Nevertheless, due to the lack of emotion based to compose a symphony, we could not get to approach the structure (\ie symphony subdivision in movements and structure inside each part).

A summary of the obtained results can be seen in Table \ref{Results}, as we have progressively studied the output generated with both approaches, with a continuous human-validation process. The system can return solo scores, but also duos, trios, quartets and an orchestra score, although we have not got to generate the score trained with all the existing symphonies due to the need of computational power. 

The human interpreter is always the source of emotions, so it is remarkable the lack of dynamics in the generated music, being played all the notes at the same volume during the whole piece. In this paper we have focused in the notes production and instruments coordination, so generated scores have not notation of the dynamics.

\section{Future work}
Following the problem exposed in the conclusion, the next step is to research in music expressiveness, in order to transmit it to the system, to obtain music similar to what a human composer would create (\ie including dynamics and musical structure and intentionality).
An option to start working in this task could be to obtain the score's dynamics, and train a Deep Learning model with the expressiveness of the work, in order to generate a \textit{template}, which would be the equivalent to the composer's way to capturing his or her feelings. After generating the dynamics (\ie the \textit{"most human"} or sentimental part) a system like the one created for this work would generate the notes and they would be fitted in the dynamic's template. 
Another improvement that could be made to the developed system is to establish more elaborated musical rules to generate notes. For instance, taking First violin's melody as the main motive, while generating other instrument's notes.

The social awareness and unconcern should be progressively made, by calming down the latent discussion around Artificial Intelligence.
In case of this paper, the most affected community are the music composers, worried of being substituted by machines. This last fact should be contradicted by clarifying that Artificial Intelligence will work as a tool to enhance their production, but, at this point, it will not generate any score without a composer's help.






\bibliographystyle{iccc}
\bibliography{iccc}

\end{document}